\documentclass[aps,pra,showpacs,reprint,superscriptaddress]{revtex4-1}
\usepackage{amsmath}
\usepackage{amssymb}
\usepackage{graphicx}
\usepackage{bbm}
\usepackage[colorlinks=true,urlcolor=blue,menucolor=blue,citecolor=blue,linkcolor=blue]{hyperref}
\usepackage{xcolor}
\usepackage{todonotes}

\newcommand{\ket}[1]{|{#1}\rangle}
\newcommand{\bra}[1]{\langle{#1}|}

\newcommand{\tr}{\mathrm{Tr}}
\makeatletter
\def\myfnsymbol#1{\expandafter\@alph2\csname c@#1\endcsname}
\def\@myfnsymbol#1{\ensuremath{\ifcase#1\or \dagger\or \ddagger\or *\or
   \mathsection\or \mathparagraph\or \|\or **\or \dagger\dagger
   \or \ddagger\ddagger \else\@ctrerr\fi}}
\let\@fnsymbol\@myfnsymbol
\makeatother
\begin{document}
\title{Quantum state tomography as a numerical optimization problem}
\author{Violeta N.~Ivanova-Rohling}
\email{violeta.ivanova-rohling@uni-konstanz.de}
\affiliation{Department of Physics, University of Konstanz, D-78457 Konstanz, Germany}
\affiliation{Zukunftskolleg, University of Konstanz, D-78457 Konstanz, Germany}
\affiliation{%
Department of Mathematical Foundations of Computer Sciences, Institute of Mathematics and Informatics, Bulgarian Academy of Sciences,
Akad.\ G.\ Bonchev, block 8,
1113 Sofia, Bulgaria}
\author{Guido Burkard}
\email{guido.burkard@uni-konstanz.de}
\affiliation{Department of Physics, University of Konstanz, D-78457 Konstanz, Germany}
\author{Niklas Rohling}
\email{niklas.rohling@uni-konstanz.de}
%\thanks{Corresponding author.}
\affiliation{Department of Physics, University of Konstanz, D-78457 Konstanz, Germany}
\begin{abstract}
We present a framework that formulates the quest for the most efficient quantum state tomography scheme as an optimization problem which can be solved numerically.
This approach can be applied to a broad spectrum of relevant setups including measurements restricted to a subsystem.
To illustrate the power of this method we present results for
the six-dimensional Hilbert space constituted by a qubit-qutrit system, which could be realized e.g.\ by the $^{14}$N nuclear spin-1 and two electronic spin states  of a nitrogen-vacancy center in diamond.
Measurements of the qubit subsystem are expressed by projectors of rank three, i.e., projectors on half-dimensional subspaces.
For systems consisting only of qubits, it was shown analytically that a set of projectors on half-dimensional subspaces can be arranged in an informationally optimal fashion for quantum state tomography, thus forming so-called mutually unbiased subspaces.
Our method goes beyond qubits-only systems and we find that in dimension six such a set of mutually-unbiased subspaces can be approximated  with a deviation irrelevant for practical applications.
\end{abstract}

\maketitle

\section{Introduction}
There has been a strong interest in quantum computing since the publication of Shor's algorithm \cite{Shor} for prime factorization. Among other tasks performed efficiently by quantum computers are quantum simulations \cite{QuantumSimulationsRMP}, aiming at finding the state of a system which is described by quantum mechanics or to compute its time evolution.
Many physical platforms have been suggested for building a quantum computer, including trapped ions \cite{ionsreview}, superconducting qubits comprising Josephson junctions \cite{Aruteetal2019,supercondqubitsreview}, electron spins in semiconductor quantum dots \cite{LossDiVincenzo,KloeffelLoss2013}, and electron or nuclear spins at a nitrogen-vacancy (NV) defect in diamond \cite{NV-centers}.
However, despite impressive results regarding the coherent control and coupling of qubits, the implementation of a general purpose quantum computer with a number of qubits relevant for practical applications remains a challenge. 

Any physical system which is supposed to function as a building block of a quantum computer would require tests of its functionality.
The measurements and computations which allow the estimation a quantum state are called quantum state tomography (QST) \cite{QSTbookchapter2004}.
Alongside methods to characterize quantum processes, such as quantum process tomography, randomized benchmarking (RB) \cite{Emersonetal2005,Mavadiaetal2018}, and gate set tomography \cite{Merkeletal2013,Blume-Kohoutetal2013,Mavadiaetal2018,Nielsenetal2020}, QST is part of the emerging field of quantum characterization, verification, and validation (QCVV), which is dedicated to the above mentioned tests of quantum systems.
QST is a central tool for verifying and debugging a quantum device
and can be helpful for the process of implemention of
a quantum computer in a physical system.
It allows for checking of the initialization of the quantum device and -- as a building block of quantum process tomography -- also the quantum gates. Therefore, the scaling of QST is not only relevant for characterizing the initialization procedure within quantum computing but also for testing quantum gates.
Specifically, quantum process tomography can be done by performing QST many times with different initial states \cite{NielsenChuang} or even by QST with one initial state by using an ancillary system \cite{Altepeteretal2003,Leung2003}.
The QST procedure calls for the acquisition of the full information of a quantum state, which requires numerous repetitions of a set of measurements and is typically very time-consuming.
Compared to other QCVV methods like low-rank tomography \cite{Flammiaetal2012} or RB, full quantum tomography is complex and comprehensive information is gained.

Because full QST is such a time consuming task, finding the optimal QST scheme, where optimal means fastest while achieving the desired precision, is thus of high practical relevance.
For a minimal set of non-degenerate measurements, this problem was considered by Wootters and Fields \cite{WoottersFields89}.
For an $n$-dimensional Hilbert space, the ideal choice is a set of $n+1$ measurement operators whose eigenbases are mutually unbiased bases (MUBs) \cite{WoottersFields89}.
Improvements to QST using MUBs are possible by allowing (i) for more than the minimum number of measurements \cite{deBurghetal2008}, (ii) for generalized measurements using ancillary systems yielding symmetric, informationally complete positive operator-valued measures (SIC-POVMs) as optimal measurements \cite{Rehaceketal2004,Renesetal}, and (iii) for adjusting the choice of measurements on the run \cite{HuszarHoulsby2012,Straupe2016,Granadeetal2017}.
Wootters and Fields \cite{WoottersFields89} introduced a geometric quality measure to evaluate the QST measurement set.
It is important to note that the use of this quality measure is not limited to non-degenerate measurements.
We have already applied this measure in the scenario where the measurements distinguish one state from the remaining $(n{-}1)$-dimensional subspace.
These measurements are described as independent rank-1 projectors.
The states can be chosen such that they belong to a set of MUBs \cite{RohlingBurkard2013}, but a numerically optimized set of measurements outperforms the MUBs \cite{VioletaNiklas}.
Furthermore, the geometric quality measure is not limited to rank-1 projection operators; on the contrary, we use it in this paper to evaluate a quorum of projection operators of higher rank.

\begin{figure}
    \centering
    \includegraphics[width=\columnwidth]{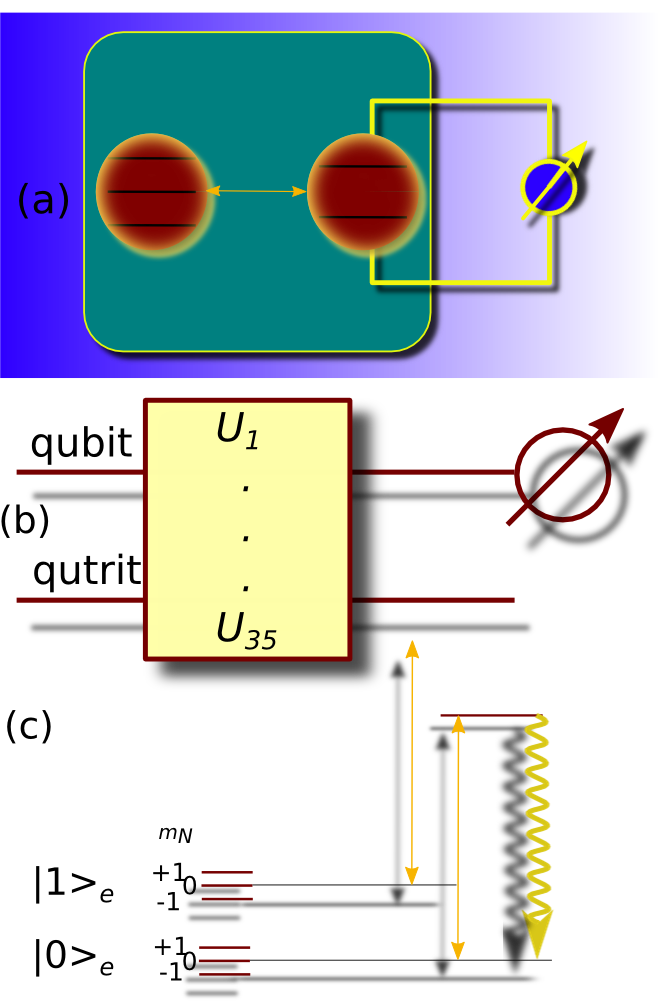}
    \caption{(a) Measurement setup for a qubit-qutrit system: only the qubit is measured, all unitary operations on the system including entangling gates between the qubit and the qutrit are available. (b) Quantum circuit of the measurement procedure: each of the 35 rank-3 projectors needed for a QST quorum are realized by applying a unitary operation $U_j$ and then performing a projective measurement on the qubit to distinguish between the qubit states $\ket{0}$ and $\ket{1}$. The resulting projection operators are then given by
$P_j=U_j \ket{0}\bra{0}\otimes\mathbbm{1}_3 U_j^\dagger$
where $\mathbbm{1}_3$ is the unity operation for the qutrit.
Note that $U_1=\mathbbm{1}$.
(c) The projection on the qubit state $\ket{0}$ can be realized for an electronic state in an NV center in diamond by resonance fluorescence where only the electronic state $\ket{0}_e$ is excited by radiation and leads to fluorescence, thus allowing for the read-out of the electronic state in the basis $\{\ket{0}_e,\ket{1}_e\}$.
The state of the nuclear spin, $m_N$, is not measured directly.}
    \label{fig:scheme}
\end{figure}

We describe a general framework to formulate  the search for an optimal QST measurement scheme as an optimization problem and use numerical methods to solve it.
To illustrate the power of this method, this paper examines the settings where only a part of a composite system is accessible to direct measurements. The relevance of this scenario becomes clear when considering the following quantum computer architecture. One logical qubit is realized by a set of physical qubits and only one of the physical qubits is equipped with a measurement device.
This can save resources on the hardware level compared to a system where each physical qubit is assigned its own measurement device.
For a quantum algorithm to be performed, reading out one physical qubit out of the set of physical qubits which constitute the logical qubit is sufficient. However, the ancilla physical qubits are needed for  quantum error correction.
We have to require that universal quantum gates are available, i.e., any unitary operation can be performed in the Hilbert space describing this quantum system.
The reasons for this are that universal quantum gates are needed for a general-purpose quantum computer as well as for realizing different measurements in the tomography scheme considered in this paper as we describe in the following.
Results for optimal QST by measuring one out of several qubits are already available \cite{BodmannHaas2018}.
Therefore, we consider here the simplest composite system which does not consist only of qubits, i.e., a qubit-qutrit system, see Fig.~\ref{fig:scheme}.
We describe the realization of such a system -- in NV centers in diamond.
We reveal the relation between our optimization problem of finding the optimal QST measurement set and packing problems in Grassmannian manifolds, which have been studied in great detail \cite{Conwayetal1996,ShorSloane1998,Calderbanketal1999,Dhillonetal2008,BodmannHaas2016,KocakNiepel2017,ZhangGe2018,Casazzaetal2018,BodmannHaas2018,Jasperetal2019,Casazzaetal2019} and are relevant for many fields, such as wireless communication, coding theory, and machine learning  \cite{Jasperetal2019,ZhengTse2002,StrohmerHeath2003,LoveHeatStrohmer2003,YapRobertsPrabhu2019}.
As we are able to approximate the optimal measurement scheme of the qubit-qutrit system, we solve a greater problem, namely we find an optimal Grassmannian packing of half-dimensional subspaces in Hilbert space of dimension six.
 
\section{Results}
\subsection{Our general framework for QST optimization}
\label{sec:general_framework}

\begin{figure*}
    \centering
    \includegraphics[width=0.8\textwidth]{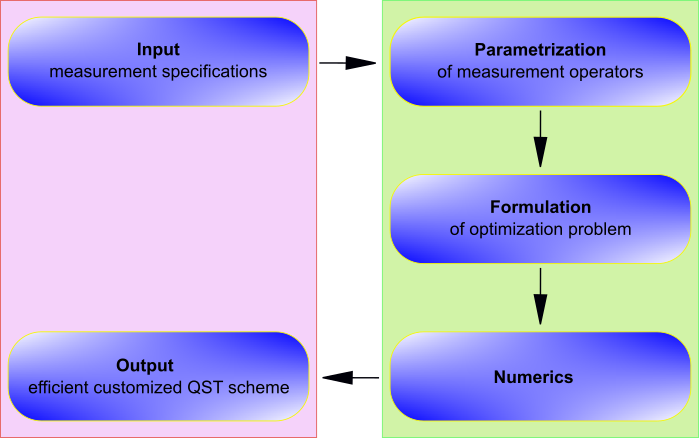}
    \caption{Workflow of the customized QST framework: On the left (red rectangle) is the user interface; on the right (green rectangle) are the internal process modules.}
    \label{fig:workflow}
\end{figure*}

Now we present our general framework for finding an optimal QST measurement scheme for a user-specified  system of finite dimension $n$, by solving a corresponding optimization problem.
In Fig.~\ref{fig:workflow}, the framework of customized QST is visualized: the user interface consisting of input measurement specifications and an efficient customized QST scheme as the output is abstracted away from the internal computational modules.
Importantly, this procedure is quite flexible and allows to include case-specific constraints to the available measurement.

The input is formed by the specifications and restrictions of the available measurements for the quantum system under investigation.
An important example of a restriction to measurement operators is the specific rank of the projectors, e.g. Ref.~\cite{VioletaNiklas} considers two-outcome measurements where the outcomes correspond to a rank-1 projector or to a rank-$(n-1)$ projector, respectively.
Importantly, the situation where only a subsystem of the quantum system is measured can be described by  a restriction on the ranks of the projection operators. Namely, the ranks of projection operators corresponding to this measurement are at least the dimension of the subsystem's complement. In this paper, we investigate in detail the restriction to measuring one qubit as part of a composite system yielding measurements described by projectors of rank $n/2$.

After the specifications of the system of interest are formulated, we parametrize the measurement operators such that the parameters determine the states in Hilbert space which span the subspace, corresponding to the projection operators. This allows us to use a minimal number of parameters for each projector and thus minimizes the dimension of the optimization problem. 

In our framework, we adopt the geometric quality measure, as defined in Wootters and Fields \cite{WoottersFields89}.
For a set of measurement operators, each measurement operator can be represented in its spectral decomposition, i.e., as the sum of its eigenvalues times the projectors onto the respective eigenspaces.
Then, the quality measure is defined as the volume spanned in operator space by the traceless parts of the projectors. 
Wootters and Fields \cite{WoottersFields89} considered the case of non-degenerate measurements where each measurement is represented by $n$ rank-1 projection operators projecting on the eigenstates of the measurement operator.
As the eigenvectors of one measurement operator form an orthogonal basis, the optimization problem lies in optimizing the relation between the different measurement operators or -- in other words -- between their eigenbases.
In the case considered in \cite{VioletaNiklas}, the optimization problem is to arrange independent rank-1 projectors maximizing the geometric quality measure.
Here, we extend the use of this quality measure to degenerate measurements where the measurement operators are denoted by projection operators which can be of rank higher than one.
This is relevant for situations where a subsystem is measured rather than the full system.
Further below, we will focus on measurements represented by projectors on half-dimensional subspaces.
However, the approach could also easily cover other cases, e.g. measuring the qutrit in a qubit-qutrit system, where each measurement is described by three rank-2 projectors, two of which are independent.
We use the geometric measure detailed above in the formulation of our optimisation problem. 
A  formal description for the specific example solved in this paper is provided below.

We then tackle this problem by numerical means.
For a system of small dimension, including the system of dimension six considered in this paper, a standard numerical method, namely Powell's derivative free method \cite{Powellmethod64}, suffices to yield good results. For larger systems, the problem of finding optimal measurement schemes calls for more sophisticated approaches, such as ones based on machine learning and deep learning.

The output of our framework is a set of measurement operations, determined by the system's specifications and restrictions which have been given as input.
This set of measurements allows the user to perform the fastest state tomography procedure possible reaching a desired precision or the most precise for a given time.

\subsection{QST by measuring a qubit in a composite system}
\label{sec:generalsetup}

Despite the limitation of measuring no more than one qubit in each run, complete state tomography is possible if we combine the available measurement with unitary transformations.
We ask what is the ideal choice of a minimal measurement set (quorum) for QST.
If one out of $N$ qubits is measured, a complete set of MUBs can be harnessed to construct an ideal quorum in the sense that the traceless parts of the rank-$2^{N-1}$ projectors form a hypercube \cite{BodmannHaas2018}.
Then the geometric quality measure reaches its upper bound.
For two qubits, QST with parity readout, a scenario equivalent to measuring one of the qubits, was proposed  \cite{Seedhouseetal2020} and implemented \cite{Leonetal2020} for spin qubits in quantum dots.

For a Hilbert space of (non-prime power) dimension six, which corresponds to the qubit-qutrit system we consider in this paper, a complete set of MUBs is not available.
The goal of this paper is to show that a quorum of projectors can come so close to the upper bound for the geometric quality measure described above that the deviation is without practical relevance for performing QST.
This is of practical importance because qubit-qutrit systems are among the experimentally studied quantum devices, as we will show below using a physical example. 
Moreover, the search for sets of subspaces that reach the orthoplex bound has been a topic of intense research in present times.
For systems of dimension $n$ where $n$ is a power of two, Bodmann and Haas \cite{BodmannHaas2018} constructed maximal sets of orthoplex-bound-achieving subspaces and showed that such a construction is possible only for subspaces of dimension $n/2$.
For dimension six, the existence of a orthoplex-bound-achieving set is unknown.
Our result implies that from our close-to-optimal QST quorum, a maximal set of subspaces  approximately achieving the orthoplex bound, can be  constructed, extending the results in \cite{BodmannHaas2018} to dimension six.
This finding will potentially trigger research on other composite dimensions such as ten or twelve.
Our approach of numerically approximating a smaller set of projection operators which reaches the geometric quality measure bound explained above and then using the method from \cite{BodmannHaas2018} to extend this set to a maximal set (approximately) achieving the orthoplex bound is promising for these higher-dimensional cases.

An example for a qubit-qutrit system is a negatively charged NV center in diamond \cite{ChildressHanson2013,Awshalometal2018}.
If the nitrogen nucleus is a ${}^{14}$N, then the nuclear spin is one, i.e., it represents a qutrit.
Two states of the electronic spin-1 of the NV center effectively constitute a qubit.
NV centers have been under intense investigation due to the long spin lifetimes of both, the nuclear spin and the electron state, and due to the possibility to perform unitary operations by microwave driving or by selectively exciting optical transitions between the energy levels of this quantum system \cite{ChildressHanson2013}.
Single-shot projective measurements on the lowest electronic state can be done by resonant excitation fluorescence \cite{Awshalometal2018,Duttetal2007,Neumannetal2010,Robledoetal2011}.
Such measurements can be described by rank-3 projectors which are considered here (see Fig. \ref{fig:scheme}).

\subsection{Packings in a Grassmannian Manifold}
\label{sec:grassmannian}
The subspaces of dimension $l$ described by the projection operators of rank $l$ form a vector space with special properties, called a \textit{Grassmannian manifold}.
We will define this notion below. 
Given an $n$-dimensional vector space $\mathbbm{V}$ over a field $\mathbbm{F}$, a Grassmannian $\operatorname{Gr}(l,\mathbbm{V})$
is the space of $l$-dimensional linear subspaces of  $\mathbbm{V}$.
Subspace packing in a Grassmannian manifold, or \textit{Grassmannian  packing}, is the problem of  maximizing the minimum pairwise distance in a set of subspaces.
We will describe below in detail how this packing problem relates to the problem of optimal QST in the setting considered here. We consider the case in which $\mathbbm{F} = \mathbbm{C}$ and $\mathbbm{V} = \mathbbm{C}^6$.
The problem of arranging a set of $m$ subspaces
$\{U_j \in \operatorname{Gr}(l,\mathbbm{F}^n),j=1,\ldots,m\}$
in an optimal, maximally spread fashion has been studied for both $\mathbbm{F}=\mathbbm{R}$ and $\mathbbm{F}=\mathbbm{C}$ \cite{Conwayetal1996,ShorSloane1998,Calderbanketal1999,Dhillonetal2008,BodmannHaas2016,KocakNiepel2017,ZhangGe2018,Casazzaetal2018,BodmannHaas2018,Jasperetal2019,Casazzaetal2019}.
Typically, optimality here refers to maximizing the minimum chordal distance
$d^2_c(P_j,P_i) = l - \tr(P_j^\dag P_i)$ where $P_j$ is the projector on the subspace $U_j$, i.e.,
$\min_{i\ne j}d^2_c(P_j,P_i)$ shall be maximal.

Here we consider a problem which is different from optimal spreading, as we are not interested in maximizing the smallest distance between the projectors on the subspaces but in the subspaces being informationally independent.
However, for the specific situation we consider, $\mathbbm{F}=\mathbbm{C}$, $l=n/2$, the optimal solutions for QST can be naturally extended to a maximum set of an optimal Grassmannian packing as we will discuss in the following.

\subsection{Optimality condition, upper bound, and consequences}
\label{sec:optimality}
We consider a Hilbert space of dimension $n$
and projection operators projecting onto subspaces of dimension $l$,
later we will specialize to $n=6$ and $l=3$.
Then, the measurements are described by rank-$l$ projection operators.
The matrix corresponding to a rank-$l$ projection operator has $l$ linearly independent columns, which define an $l$-dimensional subspace. Conversely, every $l$-dimensional subspace can be described by a projection operator of rank $l$.

A minimal state tomography set consists of $n^2-1$ of those projectors, $\{P_1,\ldots,P_{n^2{-}1}\}$. In this case, the problem of finding an optimal QST quorum is equivalent to the problem of arranging the projectors $P_j$ ($j=1,\ldots,n^2-1$) in an optimal fashion. We define the traceless parts of these operators $Q_j=P_j-l\mathbbm{1}/n$. As stated above, we evaluate the quorum by using the quality measure $\mathcal{Q}$ introduced by Wootters and Fields \cite{WoottersFields89}, defined as the volume spanned by $\{Q_1,\ldots,Q_{n^2{-}1}\}$
in the vector space of traceless $n\times n$ matrices with the scalar product $\tr(A^\dag B)$.
The length of the $Q_j$ in this vector space is fixed to
\begin{align}
\sqrt{\tr(Q_j^\dag Q_j) }
& = \sqrt{ \tr\left(\left(P_j{-}\frac{l\mathbbm{1}}{n}\right)^\dag\left(P_j{-}\frac{l\mathbbm{1}}{n}\right)\right) }
 =\sqrt{l{-}\frac{l^2}{n}},
\end{align}
and thus, the volume is fully determined by the angles between the $Q_j$.

An upper bound for the quality measure is
\begin{equation}
\label{eq:upper_bound}
    \mathcal{Q}_{ub} = \left(l(1-l/n)\right)^{(n^2-1)/2},
\end{equation} which is reached only if
$\tr(Q_j^\dag Q_i)= 0$ for all $i\ne j$.
Note that any rank-$l$ projector is available since we assume that it is possible to perform one basic measurement projecting on an $l$-dimensional subspace and that all unitary operations can be performed.
Below we describe how an upper-bound-reaching set of rank-$n/2$ projection operators relates to two other notions, namely mutually unbiased subspaces and quantum 2-designs.

\subsubsection{Mutually unbiased subspaces}
\label{sec:MUSs}
We want to see how reaching the upper bound compares to the chordal distance 
and find for $\tr(Q_j^\dag Q_i)= 0$,
\begin{equation}
d^2_c(P_j,P_i)  = \frac{l(n-l)}{n}.
\label{eq:orthoplexbound}
\end{equation}
This is the so-called orthoplex bound, which appears as an upper bound for the minimal chordal distance of projectors on 
$l$-dimensional subspaces in $\mathbbm{C}^n$ for a set of at least $n^2+1$ elements  \cite{BodmannHaas2018}.
If for two subspaces of the Hilbert space the corresponding projectors fulfill Eq.~(\ref{eq:orthoplexbound}), they are called \textit{mutually unbiased subspaces} \cite{BodmannHaasshorthistory}.
Now, we will focus on the case $l=n/2$, i.e., the problem of packing of \textit{half-dimensional subspaces}.
For this case, Bodmann and Haas \cite{BodmannHaas2018}
showed that if $n$ is a power of two, an optimal orthoplex-bound-achieving packing, maximal in terms of the number of its elements, exists. This packing consists of 
$n^2-1$ projectors $P_j$ whose corresponding $Q_j$ are pairwise orthogonal, and the projectors $P_{j+n^2-1}=\mathbbm{1}-P_j$ for $j=1,\ldots,n^2-1$.
The maximal number of elements of a set of projectors which achieves the orthoplex bound is $2(n^2-1)$, thus this maximum of elements is reached here.
In general -- not limited to the case of $n$ being a power of two -- for $l=n/2$, the bound simplifies to $\mathcal{Q}_{\mathrm{ub}} = (n/4)^{(n^2-1)/2}$ and the condition for the pairwise chordal distance becomes
\begin{equation}
d^2_c(P_j,P_i)  = \frac{n}{4}.
\end{equation}
For the qubit-qutrit system considered here $n=6$, and a quorum has $35$ elements, and with $l=n/2=3$, the quality measure's upper bound is given by $\mathcal{Q}_{ub}=(3/2)^{35/2}\approx 1206.69$.

\subsubsection{Quantum 2-designs}
The problem of optimal arrangement of projections is closely related to the notion of \textit{quantum t-designs}. Furthermore, quantum t-designs are known to be highly relevant to optimal QST measurement schemes.
For the situation of measuring a qubit in a qubit-qutrit system, we are interested in t-designs formed by projectors of higher rank -- namely rank three.
Nevertheless, we will briefly first review the case of t-designs formed by rank-1 projection operators.
In this case, quantum t-designs can be defined as  sets of projectors $\{\ket{\psi_j}\bra{\psi_j}; j=1,\ldots,N\}$ on the states $\ket{\psi_j}$ and corresponding weights $p_j>0$ with $\sum_{j=1}^Np_j=1$ which fulfill \cite{RoyScott2007}
\begin{equation}
\sum_{j=1}^N p_j \ket{\psi_j}\bra{\psi_j}^{\otimes t}
=
\int\, d\psi \ket{\psi}\bra{\psi}^{\otimes t}
\end{equation}
where the integral is taken over a uniform distribution of all states of the Hilbert space.
Positive operator-valued measures (POVMs) are 1-designs
\cite{Scott2006}.
Examples of quantum 2-designs with equal weights, $p_j=1/N$ for $j=0,\ldots,N$, are 
SIC-POVMs \cite{Renesetal} and complete sets of MUBs \cite{KlappeneckerRoetteler}.
If complete sets of MUBs are not available, as is the case for dimension six, the construction of weighted 2-designs with non-equal weights can be useful \cite{RoyScott2007}.
Under the assumption of linear reconstruction, it has been shown that quantum 2-designs are ideal for
QST performed by one repeated generalized measurement described by one informationally complete POVM
\cite{Scott2006} and for projective non-degenerate measurements \cite{RoyScott2007}.

In Refs.~\cite{Appleby, BodmannHaas2018, Zaunerthesis}, quantum t-designs of higher rank have been investigated and examples have been constructed.
Appleby \cite{Appleby} has found quantum 2-designs of higher rank which behave similarly to SIC-POVMs, termed symmetric informationally complete measurements (SIMs).
The maximal orthoplex-bound-achieving sets of half-dimensional subspaces discussed above are examples of higher-rank ($n/2$) quantum 2-designs \cite{Zaunerthesis,BodmannHaas2018}.
It was first considered by Zauner \cite{Zaunerthesis} as an example of a quantum 2-design consisting of operators of higher rank.
Bodmann and Haas \cite{BodmannHaas2018} explicitly construct these 2-designs using complete sets of MUBs and Johnson codes.

\subsection{Numerical Results}
\label{sec:results}
The best result we obtained numerically for the geometric quality measure is
\begin{equation}
    \mathcal{Q}_{\mathrm{num}} = 1206.53,
\end{equation}
which corresponds to a deviation of $\Delta\mathcal{Q}/\mathcal{Q}_{\mathrm{ub}}=(\mathcal{Q}_{\mathrm{ub}}-\mathcal{Q}_{\mathrm{num}})/\mathcal{Q}_{\mathrm{ub}}=1.3\times10^{-4}$.
For the following measure of non-orthogonality,
\begin{equation}
\label{eq:non-orthogonality}
L=\sum_{i\ne j}|\tr(Q_i^\dag Q_j)|,
\end{equation}
where $i,j\in\{1,\ldots,35\}$,
this quorum yields $\ln(L)=-0.08394$.
We include the corresponding parameters which determine the rank-3 projection operators of the quorum as well as the implementation of the computation of the quality measure $\mathcal{Q}_{\rm num}$ and of $\ln(L)$ from these parameters in Supplemental Materials available at \cite{SupplMat}.

Certainly, coming close to the upper bound for the geometric quality measure is not a proof of the existence of a quorum which actually achieves the upper bound.
However, for practical purposes, the small deviation of our numerical result from the upper bound is inconsequential for the following reasons.
The average information gain $\langle\mathcal{I}\rangle$, quality measure $\mathcal{Q}$, and number of repetitions of each of the measurements, $N_{\mathrm{rep}}$ obey the relation \cite{WoottersFields89}
\begin{equation}
    \langle\mathcal{I}\rangle = const. +\frac{n^2-1}{2}\ln\left(\frac{N_{\mathrm{rep}}}{2}\right)  + \ln(\mathcal{Q}).
\end{equation}
In our case, the additive constant differs from that in Ref.~\cite{WoottersFields89}. However, this does not affect the scaling of the required number of repetitions with the quality measure if a desired value for the average information gain must be achieved,
\begin{equation}
    N_{\mathrm{rep}} \sim\mathcal{Q}^{-2/(n^2-1)}.
\end{equation}
Here ($n=6$), the relative deviation of $1.3\times10^{-4}$ for the quality measure corresponds to a necessary relative increase in the number of repetitions of merely $10^{-5}$.
This implies that if $N_{\mathrm{rep}}=10^5$ for the ideal quorum, the deviation in quality of our quorum can be compensated by just one more repetition of each of the measurement.

\section{Discussion}
\label{sec:conclusions}
In this paper, we have optimized a QST scheme for a qubit-qutrit system where only the qubit can be measured directly and all unitary operations are available.
The quality of our solution approximates the upper bound which corresponds to the situation where the measurements project onto subspaces which are mutually unbiased.
For practical purposes, the disadvantage of not fully achieving the upper bound can be disregarded.
From a mathematical perspective, however, the explicit construction of a set of 35 mutually unbiased three-dimensional subspaces in $\mathbbm{C}^6$ remains an open problem.
Such a construction might also allow a generalization to higher composite dimensions such as ten or twelve, where the numerical approach is significantly more difficult than for the six-dimensional case studied in this paper.
While this example of a qubit-qutrit system is of importance in its own right given its realization by an NV center, our general approach can be applied to a broad range of QST problems under limited measurements.
This might allow experimentalists to find the most optimal QST scheme for their specific system.

Our method of numerically solving the smaller-dimensional problem of finding a set of projection operators, optimal for QST in the sense of \cite{WoottersFields89}, and then extending this set to build a maximal set which approximates the orthoplex bound may be employed for looking for approximations of maximal orthoplex-bound-achieving sets in higher dimensions.

For a higher dimension $d>6$, the optimization problem becomes computationally more challenging.
Further future research might include the application and tailoring of machine learning methods to the  high-dimensional optimization problem.
In Ref.~\cite{VioletaNiklasCIT2020}, we have already applied machine learning methods and obtained rank-1 QST quorums in dimension eight which are improved compared to the result achieved by standard numerical methods used in \cite{VioletaNiklas}.

\section{Methods}
\subsection{Optimization problem}
\label{sec:optimizationproblem}
The QST quorum for a qubit-qutrit system consists of 35 measurements each described by a projector on a three-dimensional subspace of the six-dimensional Hilbert space.
In order to parametrize the projectors, we use three pairwise orthogonal vectors of the Hilbert space.
By vectors we mean here normalized vectors with arbitrary global phase.
In general, such a vector in $\mathbbm{C}^6$ is given by ten real parameters.
However, we can choose each of the three vectors effectively in a four-dimensional Hilbert space.
The reason is the dimensionality of the involved spaces: for any three-dimensional subspace and any four-dimensional subspace of a six-dimensional Hilbert space there is at least one vector which is a common element of both subspaces.
If we have chosen the first vector of our three-dimensional subspace in this way, we can choose the second vector from a four-dimensional subspace of the five-dimensional space which is orthogonal on the first vector.
Analogously, any two-dimensional subspace and any four-dimensional subspace of a five-dimensional Hilbert space have at least one vector in common.
Finally, the third vector is chosen from the remaining four-dimensional subspace orthogonal on the first and the second vector.
Each of the vectors, denoted in a basis of the respective four-dimensional subspace by $\ket{\psi}=\sum_{i=1}^4 x_i \ket{i}$, is given by six real parameters, $\theta_1,\theta_2,\theta_3$ and $\varphi_2,\varphi_3,\varphi_4$, in the following way,
\begin{align}
x_1 & = \cos\theta_1,\\
x_2 & = \sin\theta_1\cos\theta_2 e^{i\varphi_2},\\
x_3 & = \sin\theta_1\sin\theta_2\cos\theta_3e^{i\varphi_3},\\
x_4 & = \sin\theta_1\sin\theta_2\sin\theta_3 e^{i\varphi_4}.
\end{align}
We compute a unitary operation which maps the second vector into the space orthogonal to the first.
Then we compute a unitary operation which maps the third vector on the space orthogonal to the first and the second vector.
Thus, each projector is given by 18 real parameters.
Furthermore, we know that the quorum performance is invariant under any unitary operation on the Hilbert space.
Therefore, we can choose for the first projector without loss of generality, $P_1=\operatorname{diag}(1,1,1,0,0,0)$, i.e.,
that it projects on the first three basis states for whatever basis we have chosen.
Overall, our optimization problem has $N_{\rm params}=34\times3\times6=612$ real parameters.

\subsection{Numerics}
\label{sec:numerics}
As in \cite{VioletaNiklas}, we apply Powell's derivative free method to numerically optimize the set of measurement operators.
Coming close to the upper bound for the quality measure, $\mathcal{Q}_{ub}=(3/2)^{35/2}$, see Eq.~(\ref{eq:upper_bound}),
we conjecture that $\mathcal{Q}_{ub}$ can indeed be reached.
In the following we can make use of this conjecture because then a quorum which reaches the maximum for the geometric quality measure also has no non-orthogonal contributions for the matrices $Q_1,\ldots,Q_{35}$.
As the Powell method for maximizing the volume $\mathcal{Q}$ in operator space converges slowly,
we additionally consider the quantity $L$ defined in Eq.~(\ref{eq:non-orthogonality})
which is a measure for the non-orthogonality of the $Q_1,\ldots,Q_{35}$
and alternating
with maximizing the volume, we aim to minimize $\ln(L)$ again with Powell's method.

\section*{Data availability}
The parameters for the best quorum we have found as well as a python program which computes the geometric quality measure $\mathcal{Q}$ and the logarithm of the non-orthogonality measure, $\ln(L)$ are available online \cite{SupplMat}.

\section*{Acknowldgements}
This work was partially supported by the Zukunftskolleg (University of Konstanz) and the Bulgarian National Science Fund under the contract No KP-06-PM 32/8.

\section*{Author contributions}
VNI-R and NR developed the idea of customized QST and implemented the numerics for the rank-3 projectors in dimension six.
GB identified and discussed the example for a qubit-qutrit system.
All authors participated in the discussion of the results and in writing the manuscript.

\end{document}